\documentclass{jkas}

\def\year{2015} 
\def\month{} 
\def\volume{00} 
\def\issue{0} 
\def\beginpage{1} 
\def\endpage{11} 
\def\received{XXX, 2015} 
\def\accepted{XXX, 2015} 
\date{Received \received ; accepted \accepted}

\title{Photometric study of IC 2156}

\author[]{A. L. Tadross}
\author[]{Y. H. M. Hendy}

\affil[]{National Research Institute of Astronomy and Geophysics, 11421 - Helwan, Cairo, Egypt; \email{altadross@yahoo.com}}

\def\corrauthor{
A. L. Tadross
}

\def\runningauthor{
A. L. Tadross
}

\def\runningtitle{
Photometric study of IC 2156
}

\def\keywords{
Galaxy: open clusters and associations - individual: IC 2156 - astrometry - Stars: luminosity function - Mass function.
}

\def\abstracttext{
The optical {\it UBVRI} photometric analysis has been established using SLOAN DIGITAL SKY SURVEY (SDSS database) in order to estimate the astrophysical parameters of poorly studied open star cluster IC 2156. The results of the present study are compared with a previous one of ours, which relied on the 2MASS {\it JHK} infrared photometry. The stellar density distributions and color-magnitude diagrams of the cluster are used to determine the geometrical structure; limited radius, core and tidal radii, the distances from the Sun, from the Galactic plane and from the Galactic center. Also, the main photometric parameters; age, color excesses, reddening-free distance modulus, membership, total mass, luminosity function, mass function and relaxation time; have been estimated.
}

\begin{document}
\jkashead 


\section{Introduction}
Open star clusters are important celestial bodies in understanding star formation and stellar evolution theories. Color-magnitude Diagram (CMD) analysis through isochrones gives us good estimates of the astrophysical parameters of the clusters, e.g. age, reddening and distance. In the last decades, many studies have been performed using different techniques; started from photographic photometry to the charge-coupled device (CCD) photometry, and finally employing many isochrones models. The large amount of results, which produced in the literature are gathered in catalogs and databases, e.g. Webda\footnote{http://www.univie.ac.at/webda/navigation.html} and Dias\footnote{http://www.wilton.unifei.edu.br/ocdb/}. In this context, we have presented some contributions of papers (Tadross 2011; Tadross 2009a; Tadross 2009b; Tadross 2008a; Tadross 2008b).

The present study depends mainly on the last version of SDSS database (SDSS DR12)\footnote{http://www.sdss.org/dr12/}, which provides homogeneous {\it ugriz} photometry for stars in the northern sky. The most important reason for using SDSS database lies in the {\it ugriz} point-spread function (PSF) photometry for setting the zero points of {\it UBVRI}, Chonis \& Gaskell (2008). The cluster IC 2156 was studied in JHK 2MASS system by Tadross (2009b) among 11 previously unstudied open star clusters. Fig. 1 displays the image of this target.
\\ \\
This paper is organized as follows. Data extraction is presented in Section 2, while the data analysis and parameter estimations are described in Sections 3. Finally, the results and conclusion of our study are summarized in Section 4.

\begin{figure*}
\begin{center}
      {\includegraphics[width=8cm]{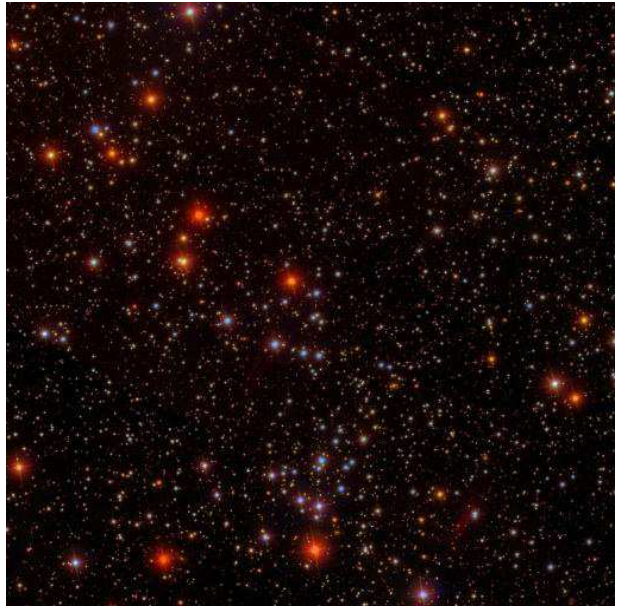}}
      \end{center}
      \caption{The image of IC 2156, as taken from the SDSS website$^{3}$.}
\end{figure*}

\section{Data Extraction}

The open star cluster IC 2156 located at J2000.0 coordinates $\alpha=06^{h} \ 04^{m} \ 51^{s}, \ \delta= +24^{\circ} \ 09^{'} \ 30^{''}, \ \ell= 186.291^{\circ}, \ b= 1.297^{\circ}$. We extracted {\it ugriz} PSF magnitude of all stars around the center of the cluster in a radius of 10 arcmin from the SDSS data release 12 (DR12) by Alam et al. (2015). The SDSS survey conducted with the CCD of the 2.5 m telescope at Apache Point (New Mexico, USA). We had to convert {\it ugriz} magnitudes into the {\it UBVRI} photometric system (Johnson-Cousins) using Chonis \& Gaskell (2008). The standard errors of the transformation equations for {\it U}, {\it B}, {\it V}, {\it R} and {\it I} are 0.007, 0.007, 0.005, 0.005 and 0.009 respectively. Fig. 2 represents the magnitude errors in each filter {\it ugriz}.
\\ \\
Although the apparent diameter of the cluster is less than 5 arcmin, the downloaded data are taken to be exceeded that diameter, i.e. about 10 arcmin because it should be reached the background field stars. To get net worksheet data for the investigating cluster, the photometric completeness limit has been applied to the photometric pass-band SDSS data to avoid over-sampling of the lower parts of the cluster's CMDs (cf. Bonatto et al. 2004). Stars with observational uncertainties $\geq$ 0.20 mag have been removed. In addition, stellar photometric membership criteria are adopted based on the location of the stars within $\pm$ 0.1 mag around the zero age main sequence (ZAMS) curves in the CMDs (Clari$\acute{a}$ \& Lapasset 1986).

\begin{figure*}
\begin{center}
      {\includegraphics[width=15cm]{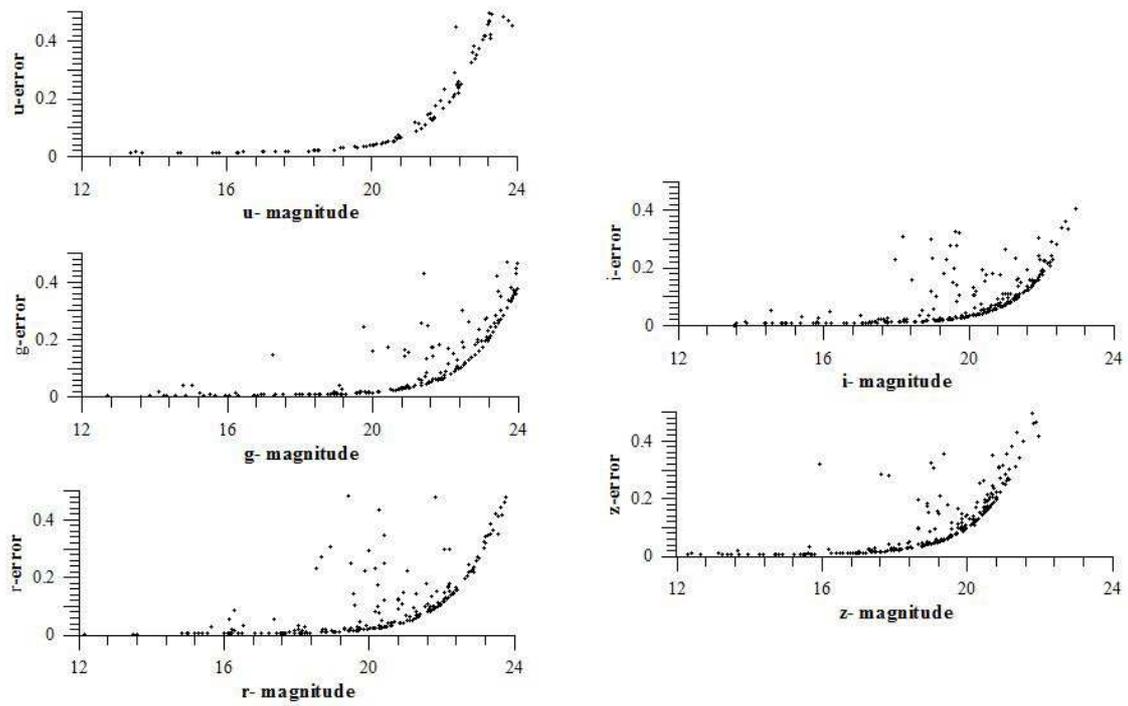}}
      \end{center}
      \caption{The magnitudes errors of the cluster IC 2156.}
\end{figure*}

\section{Data Analysis}
\subsection{Cluster's Radial Density Profile}

To establish the radial density profile (RDP) of IC 2156, the area is divided into central concentric circles with bin sizes R$_{i}$ $\leq$ 1 arcmin, from the cluster center. The number density, R$_{i}$, in the i$^{th}$ zone is calculated by using the formula of R$_{i}$ = N$_{i}$/A$_{i}$ where N$_{i}$ is the number of stars and A$_{i}$ is the area of the i$^{th}$ shell. The star counts of the next steps should be subtracted from the previous ones, so that we obtained only the amount of the stars within the relevant shell's area, not a cumulative count. The density uncertainties in each shell were calculated using the relative error of Poisson. We applied the empirical King model (1966), parameterizing the density function $\rho(r)$ as:

\begin{center}
{\Large $\rho(r)=f_{bg}+\frac{f_{0}}{1+(r/r_{c})^{2}}$}
\end{center}

where $f_{bg}$, $f_{0}$ and $r_{c}$ are background, central star density and the core radius of the cluster respectively. The cluster's limiting radius can be defined at that radius which covers the entire cluster area and reaches enough stability with the background field density. Because of strong field stars contamination, it is not possible to completely separate all field stars from cluster members. The limiting radius of the cluster can be described with an observational border, which depends on the spatial distribution of stars in the cluster and the density of the membership and the degree of field-star contamination.
Fig. 3 shows the RDP of the cluster IC 2156, the limited radius, core radius and the background field density are shown in the figure. Finally, knowing the cluster's total mass (Sec. 3.4), the tidal radius can be calculated by applying the equation of Jeffries et al. (2001):

\begin{center}
\Large $R_{t} = 1.46 ~ (M_{c})^{1/3}     $
\end{center}
where $R_{t}$ and $M_{c}$ are the tidal radius and total mass of the cluster respectively.
\begin{figure}
\begin{center}
      {\includegraphics[width=8cm]{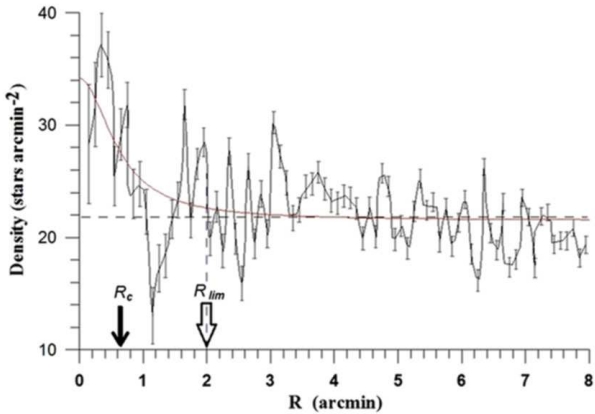}}
      \end{center}
      \caption{The radial density profile of the cluster IC 2156. Clusters' limiting radius, core radius, and the background field density are shown in the figure. The curved solid line represents the fitting of King model. Error bars are determined from sampling statistics [$1/(N)^{0.5}$ where N is the number of stars used in the density estimation at that point].}
\end{figure}

\subsection{The main photometry}

Here, we determined the main astrophysical parameters of the cluster, i.e. color excesses, age and distance modulus of IC 2156. Firstly we used some reliable data of stars with high membership probability (i.e. stars with good precision and located very close to the cluster's center) in order to derive the reddening value from the Color-Color, (U-B)-(B-V), diagram. The color excesses are found to be E(B-V)= 0.55 mag and E(B-U)= 0.39 mag, as shown in the upper panel of Fig. 4.

Secondly, we determined the age and distance modulus of the cluster by fitting isochrones to the Color-magnitude diagrams CMDs of the cluster. Several fittings on the CMDs of the cluster have been applied using the stellar evolution models of Girardi et al. (2010) of Padova isochrones, as shown in the lower panel of Fig. 4.

It is worth mentioning that the assumptions of solar metallicity are quite adequate for young and intermediate age open clusters, which are close to the Galactic disk. However, for a specific age isochrones, the fit is obtained at the same distance modulus (12.30 mag) and the same age (250 Myr) for all the diagrams V-(B-V), V-(V-I), V-(V-R), and V-(R-I). The color excesses are found obeyed Fiorucci \& Munari (2003)'s relations for normal interstellar medium as shown in Fig. 4 (lower panel). They are found to be 0.55, 0.70, 0.32, and 0.44 mag respectively.
\\ \\
Under the assumption of $R_{gc_{\odot}}$= 8.34 $\pm$ 0.16 kpc of Reid et al. (2014), which is based on high precision measurements of the Milky Way, the distance from the Galactic center $R_{gc}$ is estimated for IC 2156 and found to be 11.20 kpc. Also, the projected distances on the Galactic plane from the Sun ($X_{\odot}~\&~Y_{\odot})$ and the distance from the Galactic plane ($Z_{\odot}$), are determined to be 2865, --315, and 65 pc respectively, see Table 1. For more details about the geometry Galactic distances calculations, see Tadross (2011).

\begin{figure*}
\begin{center}
      {\includegraphics[width=11cm]{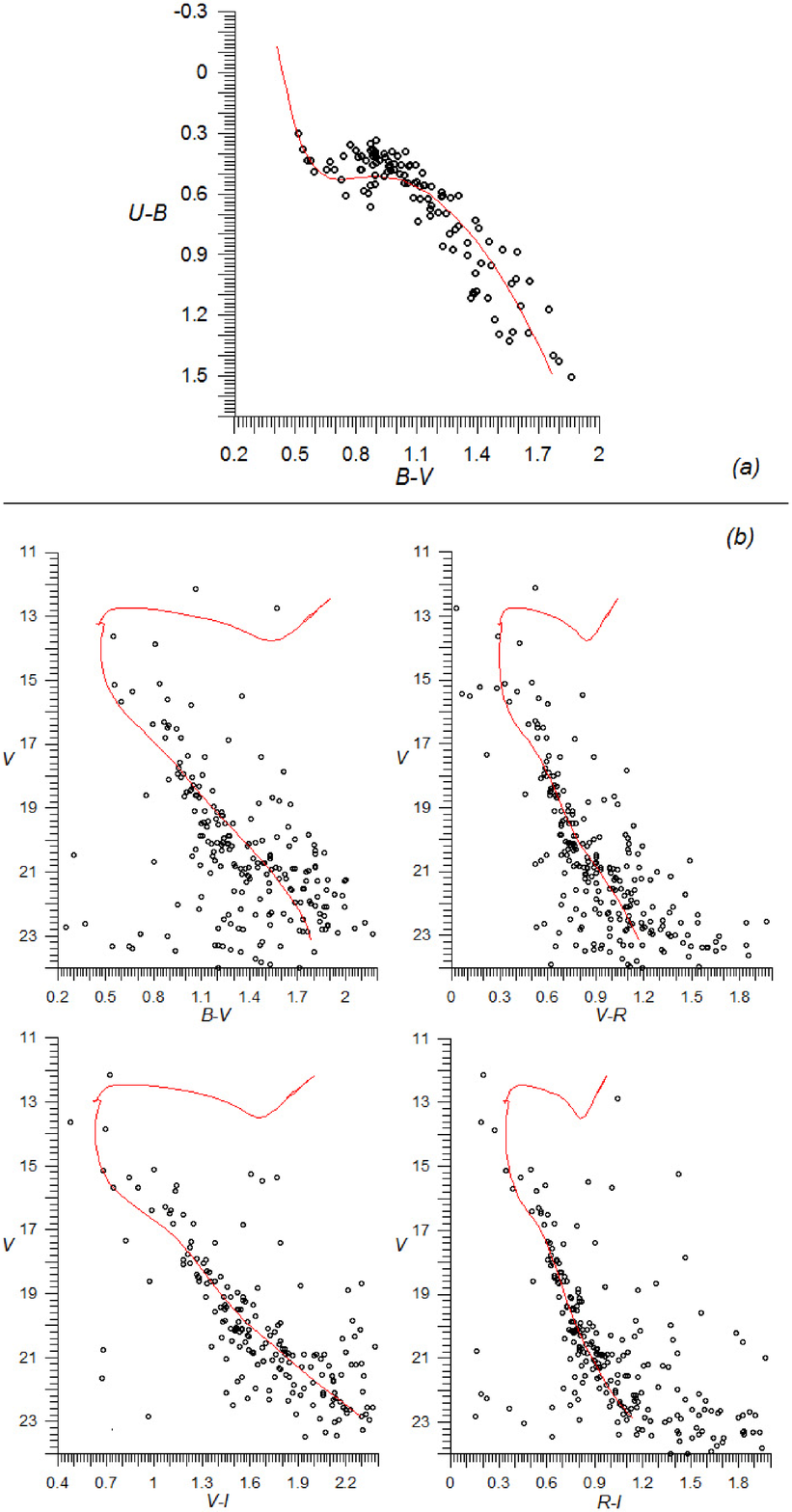}}
      \end{center}
      \caption{(a) The upper panel represents the CC-Diagram of the reliable membership probability of IC 2156, where E(B-V)= 0.55 mag and E(U-B)= 0.39 mag. (b) The lower panel represents the CM-Diagrams of the cluster, where the reddening-free distance modulus is taken at 12.30 mag, and the color excesses; E(B-V), E(V-R), E(V-I) and E(R-I); are taken to be 0.55, 0.32, 0.70 and 0.44 mag respectively.}
\end{figure*}

\subsection{Luminosity functions}

It is difficult to determine the membership of a cluster using only the stellar RDP. It might be claimed that most of the stars in the inner concentric rings are quite likely members, whereas the external rings are more intensely contaminated by field stars. Therefore, the stars, which are  closed to the cluster's center and near to the main-sequence (MS) in CMDs are taken to be the stellar membership of the clusters. These MS stars are very important in determining the luminosity, mass functions and the total mass of the investigated cluster.
For this purpose, we obtained the Luminosity Functions (LF) of the cluster by summing up the V band luminosities of all stars within the determined limiting area of the cluster. Before building the LF, we converted the apparent V band magnitudes of the cluster members into the absolute magnitude value using the distance modulus of the cluster. We constructed the histogram of LF to include a reasonable number of stars in each absolute V magnitude bins for the best counting statistics; see Fig. 5.

\subsection{Mass function and total mass}

The mass functions (MF) of the cluster is built using the theoretical evolutionary tracks and their isochrones at the specific age of the cluster. The masses of the cluster members can be derived from the polynomial expression developed by Girardi et al. (2010) with solar metallicity.

The LF and MF are correlated to each other according to the known Mass-luminosity relation. The accurate determination of both of them (LF \& MF) suffers from the field star contamination, membership uncertainty, and mass segregation, which may affect even poorly populated, relatively young clusters (Scalo 1998). On the other hand, the properties and evolution of a star are closely related to its mass, so the determination of the initial mass function (IMF) is needed. IMF is an empirical relation that describes the mass distribution of a population of stars in terms of their theoretical initial mass. The IMF is defined in terms of a power law as follows:

\begin{center}
{\Large $\frac{dN}{dM} \propto M^{-\alpha}$}
\end{center}

where $\frac{dN}{dM}$ is the number of stars of mass interval (M:M+dM), and $\alpha$ is a dimensionless exponent. The IMF for massive stars ($>$ 1 $M_{\odot}$) has been studied and well established by Salpeter (1955), where $\alpha$ = 2.35. This form of Salpeter shows that the number of stars in each mass range decreases rapidly with increasing mass. It is noted that the investigated MF slope ranging of  IC 2156 consideration is found to be -2.7, which is somewhat around the Salpeter's value as shown in Fig. 6.
\\ \\
To estimate the total mass of the cluster, the mass of each star has been estimated from a polynomial equation developed from the data of the solar metallicity isochrones (absolute magnitudes versus actual masses) at the age of the cluster. The sum of products of the number of stars in each bin by the mean mass of that bin yields the total mass of the cluster, which is found to be 310 $M_{\odot}$.

\begin{figure}
\begin{center}
      {\includegraphics[width=8cm]{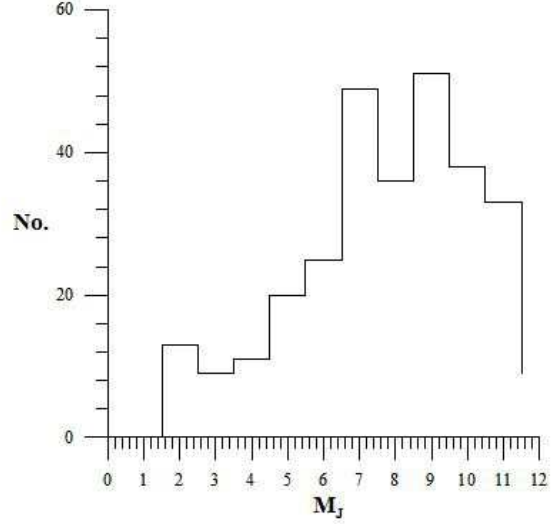}}
      \end{center}
      \caption{The luminosity functions of the cluster IC 2156.}
\end{figure}

\begin{figure}
\begin{center}
      {\includegraphics[width=8cm]{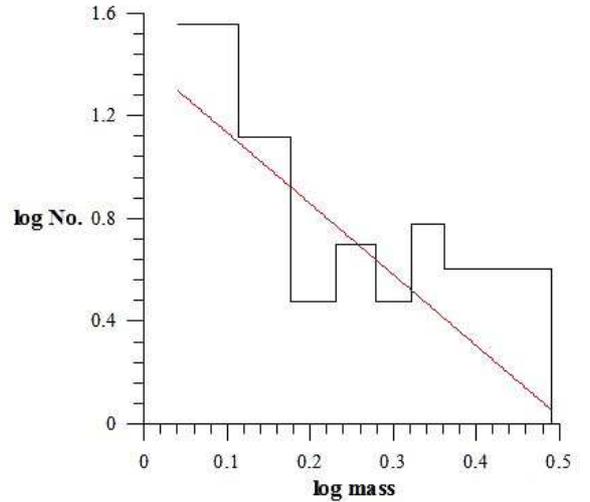}}
      \end{center}
      \caption{The mass functions of the cluster IC 2156.}
\end{figure}

\subsection{Dynamical state and relaxation time}

The time in which the cluster needs from the very beginning to build itself and reach the stability state against the contraction and destruction forces is known as the relaxation time of the cluster  ($T_{relax}$). This time is depending mainly on the number of members and the cluster diameter. To describe the dynamical state of the cluster, the relaxation time can be calculated in the form:
\begin{center}
$ T_{relax}=\frac{N}{8\ln N}~T_{cross}$
\end{center}

where $T_{cross}=D/\sigma_{V}$ denotes the crossing time, $N$ is the total number of stars in the investigated region of diameter $D$, and $\sigma_{V}$ is the velocity dispersion (Binney \& Tremaine  1998) with a typical value of 3 km s$^{-1}$ (Binney \& Merrifield 1987). Using the above formula we estimated the dynamical relaxation time for IC 2156, which found to be 5.5 Myr. It means that IC 2156 is indeed dynamically relaxed.

\section{Conclusion}

The open star cluster IC 2156 has been extracted using {\it ugriz} filter of the SDSS survey, converted to {\it UBVRI} using transformation equations of Chonis \& Gaskell (2008). This open cluster has been studied before using {\it JHK} pass-band of 2MASS database. We compared our astrophysical parameters of the cluster with Tadross (2009b), the results are summarized and listed in Table 1. Some differences have been occurred at the distances of the cluster from the Sun; from the Galactic center $R_{gc}$; from the Sun's projection location on the Galactic plane ($X_{\odot}~\&~Y_{\odot})$ and from the Galactic plane ($Z_{\odot}$) as well.

\begin{table}
\caption{Comparisons between Tadross (2009b) and the present one.}
\begin{tabular}{lll}
\hline\noalign{\smallskip}Parameter&Tadross (2009b)&The present study
\\\hline\noalign{\smallskip}
Membership&---&295 stars \\
Age&250 Myr.&250 Myr.\\
E (B--V)&0.67 mag.&0.55 mag.\\
Metal abundance (Z)&0.019&0.019\\
$(V-M_{v})_{o}$&12.20 mag.&12.30 mag.\\
Distance&2750 $\pm$ 125 pc.&2880 $\pm$ 133 pc.\\
Radius&2.0$^{'}$&2.0$^{'}$ (1.67 pc.)\\
Core radius&---&0.62$^{'}$ \\
Tidal radius&---&9.9 pc.\\
$R_g$&10.6 kpc.&11.2 kpc.\\
X$_{\odot}$&2087 pc.&2865 pc.\\
Y$_{\odot}$&--230 pc.&--315 pc.\\
Z$_{\odot}$&47 pc.&65 pc.\\
Luminosity function&---&Estimated\\
{\it IMF} slope&---&--2.7$ \pm$ 0.09\\
Total mass&---&$\approx$ 310 M$_{\odot}$\\
Relaxation time& ---&5.5 Myr.\\
\hline{\smallskip}
\end{tabular}
\end{table}

\acknowledgments

This paper regards as a part of the project No. STDF-1335; funded by Science \& Technology Development Fund (STDF) under the Egyptian Ministry for Scientific Research. The authors would like to thank the anonymous referee for his/her considerable contributions to improve the paper.
This research has made use of the SDSS DR12 (Sloan Digital Sky Survey SDSS DR12) and the Two Micron All Sky Survey (2MASS).



\begin{thebibliography}{}   	

\bibitem[Alam (2015)]{Al15} Alam, S., Albareti, F.D., Allende, P.C., et al. 2015, The Eleventh and Twelfth Data Releases of the Sloan Digital Sky Survey: Final Data from SDSS-III, ApJS 219, 12
\bibitem[Binney (1987)]{bi87} Binney, J., \& Tremaine, S. 1987, in Galactic Dynamics, Princeton series in astrophysics, Princeton University Press
\bibitem[Binney (1998)]{bi98} Binney, J., \& Merrifield, M. 1998, in Galactic Astronomy, Princeton series in astrophysics, Princeton University Press
\bibitem[Bonato (2003)]{Bon3} Bonatto, Ch., Bica, E. 2003, Mass segregation in M 67 with 2MASS, A\&A, 405, 525
\bibitem[Bonato (2004)]{Bon4} Bonatto, Ch., Bica, E., Girardi, L. 2004, Theoretical isochrones compared to 2MASS observations: Open clusters at nearly solar metallicity, A\&A, 415, 571
\bibitem[Bonato (2006)]{Bon6} Bonatto, C., Bica, E. 2006, Methods for improving open cluster fundamental parameters applied to M 52 and NGC 3960, A\&A, 455,
\bibitem[Carpen (2001)]{Car1} Carpenter, John, M. 2001, Color Transformations for the 2MASS Second Incremental Data Release, AJ, 121, 2851
\bibitem[Chonis (2008)]{Cho8} Chonis, Taylor S. \& Gaskell, C. Martin, 2008, Setting {\it UBVRI} Photometric Zero-Points using Sloan Digital Sky Survey {\it UGRIZ} Magnitudes, AJ, 135, 264
\bibitem[Claria (1986)]{Clar86} Clari$\acute{a}$, J. J., Lapasset, E. 1986, Fundamental parameters of the open cluster NGC 2567, AJ, 91, 326
\bibitem[Fiorucc (2003)]{Fio3} Fiorucci, M., Munari, U. 2003, The Asiago Database on Photometric Systems (ADPS). II. Band and reddening parameters, A\&A, 401, 781
\bibitem[Girardi (2010)]{Gira10} Girardi, L., Williams, B. F., Gilbert, K. M., Rosenfield, P., Dalcanton, J., Marigo, P., Boyer, M. L., Dolphin, A., Weisz, D. R., Melbourne, J., et al. 2010, The ACS Nearby Galaxy Survey Treasury. IX. Constraining Asymptotic Giant Branch Evolution with Old Metal-poor Galaxies, ApJ, 724, 1030
\bibitem[Jeffri (2001)]{Jef1} Jeffries, R. D., Thurston, M. R., Hambly, N. C. 2001, Photometry and membership for low mass stars in the young open cluster NGC 2516, A\&A, 375, 863
\bibitem[King (1966)]{Kin66} King, Ivan R. 1966, The structure of star clusters. III. Some simple dynamical models, AJ, 71, 64
\bibitem[Reid (2014)]{Re14} Reid, M. J., Menten, K. M., Brunthaler, A., et al. 2014, Trigonometric Parallaxes of High Mass Star Forming Regions: the Structure and Kinematics of the Milky Way, ApJ, 783, 130
\bibitem[Salpter (1955)]{Sal55} Salpeter, Edwin E. 1955, The Luminosity Function and Stellar Evolution, ApJ, 121, 161
\bibitem[Scalo (1998)]{Sca98} Scalo, J. 1998, The IMF Revisited: A Case for Variations, ASPC, 142, 201
\bibitem[Tadross (2009a)]{Tad9a} Tadross, A. L. 2009a, The fundamental parameters of 60 unstudied open star clusters (Czernik, Dol-Dzim, Kronberger, and Turner), Ap\&SS, 323, 383
\bibitem[Tadross (2009b)]{Tad9b} Tadross, A. L. 2009b, An investigation of 11 previously unstudied open star clusters, New Astronomy, 14, 200
\bibitem[Tadross (2008a)]{Tad8a} Tadross, A. L. 2008a, A Catalogue of previously unstudied Berkeley clusters, MNRAS, 389, 285
\bibitem[Tadross (2008a)]{Tad8b} Tadross, A. L. 2008b, The main parameters of 25 un-studied open star clusters, New Astronomy, 13, 370
\bibitem[Tadross (2011)]{Tad11} Tadross, A. L. 2011, A Catalog of 120 NGC Open Star Clusters, JKAS, 44, 1

\end{thebibliography}
\end{document}